\documentclass[%
 amsmath,amssymb,
 aps,
]{revtex4-2}

\usepackage{graphicx}
\usepackage{dcolumn}
\usepackage{bm}
\usepackage{xcolor}
\usepackage{listings}
\usepackage{mathtools}

\newcommand{\mtq}{\texttt{mat2qubit}}
\renewcommand{\a}{\alpha}
\renewcommand{\b}{\beta}
\newcommand{\expp}[1]{^{(#1)}}
\newcommand{\ttt}[1]{\texttt{#1}}
\newcommand{\supr}[1]{^{(#1)}}

\newcommand{\la}{\langle}
\newcommand{\ra}{\rangle}

\DeclarePairedDelimiter{\ceil}{\lceil}{\rceil}

\definecolor{codegreen}{rgb}{0,0.6,0}
\definecolor{codegray}{rgb}{0.5,0.5,0.5}
\definecolor{codepurple}{rgb}{0.58,0,0.82}
\definecolor{backcolour}{rgb}{0.95,0.95,0.92}

\lstdefinestyle{mystyle}{
    backgroundcolor=\color{backcolour},   
    commentstyle=\color{codegreen},
    keywordstyle=\color{magenta},
    numberstyle=\tiny\color{codegray},
    stringstyle=\color{codepurple},
    basicstyle=\ttfamily\footnotesize,
    breakatwhitespace=false,         
    breaklines=true,                 
    captionpos=b,                    
    keepspaces=true,                 
    numbers=left,                    
    numbersep=5pt,                  
    showspaces=false,                
    showstringspaces=false,
    showtabs=false,                  
    tabsize=2,
    language=python
}

\begin{document}


\title{mat2qubit: A lightweight pythonic package for qubit encodings of vibrational, bosonic, graph coloring, routing, scheduling, and general matrix problems}

\author{Nicolas P. D. Sawaya}
\affiliation{Intel Labs, Santa Clara, California}
\email{nicolas.sawaya@intel.com}

\date{\today}

\begin{abstract}
Preparing problems for execution on quantum computers can require many compilation steps. Automated compilation software is useful not only for easier and faster problem execution, but also for facilitating the comparison between different algorithmic choices. 
Here we describe mat2qubit, a Python package for encoding several classes of classical and quantum problems into qubit representations. 
It is intended for use especially on Hamiltonians and functions defined over variables (\textit{e.g.} particles) with cardinality larger than 2. More specifically, mat2qubit may be used to compile bosonic, phononic/vibrational, and spin-$s$ problems, as well as classical problems such as graph coloring, routing, scheduling, and classical linear algebra more generally. 
In order to facilitate numerical analyses and ease of programmability, a built-in computer algebra system (CAS) allows for fully symbolic preparation and manipulation of problems (with symbolic operators, symbolic coefficients, and symbolic particle labels) before the final compilation into qubits is performed. 
We expect this code to be useful in the preparation and analysis of various classes of physics, chemistry, materials, and optimization problems for execution on digital quantum computers.
\end{abstract}

\maketitle


\section{Introduction}



When using a qubit-based quantum computer, it is necessary to map problem instances and other primitives to a Pauli representation. This is true for both quantum Hamiltonians and classical cost functions. Some problems classes are trivial to map to qubits, for example quantum spin models (since distinguishable spin-half particles are the same as qubits), and classical optimization problems defined over binary variables (\textit{e.g.} MaxCut). Other problems require more sophisticated encoding procedures; for instance, fermionic problems such as the electronic structure problem in chemistry require the Jordan-Wigner or other encodings \cite{ortiz2001_jw,bk2002_fermionic,seeley2012_bk,cao2019chemrev,mcardle2020_rev}. 

Notably, there are many scientifically and industrially relevant problems that are \textit{not} most naturally defined over qubit, fermionic, or classical binary degrees of freedom. Such problems, for which the subsystems often have cardinality greater than 2, include bosonic \cite{somma2005manybody,liu2021boson}, vibrational/phononic \cite{mcardle2019vibr,sawaya2021infrared,ollitrault2020ansatz}, 
spin-$s$ \cite{lora2016spins}, 
graph coloring \cite{Oh19_multicoloring,wang20_xymixers,Tabi20_color,fuchs20_maxkcut}, 
routing \cite{glos2022qaoa,Salehi21_tsp}, 
and scheduling \cite{crispin13_vehicle,venturelli15_jobshop,tran16_schedaaai,Krzysztof20_railway,Dalyac21_EVs,amaro21_jobshop} problems, as well as more general classical linear algebra problems. \texttt{mat2qubit} is a Pythonic software package for encoding these Hamiltonians and cost functions into qubit operators; previously, software had been developed mainly for compiling fermionic problems \cite{openfermion,pennylane,qiskit2019} and classical problems over binary variables \cite{shaydulin2021qaoakit,cirq,pennylane,qiskit2019}.

The task of interest is to perform mappings of the following class:
\begin{equation}\label{eq:general_mapping}
\sum_i b_i M\expp{i}_0 \otimes M\expp{i}_1 \otimes \cdots \rightarrow \sum_j c_j \bigotimes \{I,\sigma_x,\sigma_y,\sigma_z\}
\end{equation}
where $M\expp{i}_\a$ are square matrix operators of arbitrary size, $I$ is the identity, and $\{\sigma_x,\sigma_y,\sigma_z\}$ are the Pauli matrices. Each subsystem on the left-hand side maps to a disjoint set of qubits; the compilation procedure is described in detail in previous work \cite{sawaya2020dlev,sawaya2022dqir}. The current version of \mtq ~allows only for trivial (equivalent to bosonic) commutation relations, \textit{i.e.} $[M_\a\expp{i},M_\b\expp{i}]=0$. This requirement is obeyed by the wide range of quantum and classical problems discussed in this work, but excludes problems such as many-body Hamiltonians over fermions \cite{ortiz2001_jw}, anyons \cite{nayak08anyon}, or parafermions \cite{fendley2012parafermionic}.


The mapping of equation \eqref{eq:general_mapping} is not unique; for each subsystem one may in principle choose any integer-to-bitstring mapping, some of which require more space (qubits) than others. One purpose of this code is to simplify and automate the implementation of a wide range of encodings. Several previous works have studied the resource tradeoffs (for example depth-space tradeoffs) between different encodings \cite{sawaya2020dlev,sawaya2020connec,hadfield2019qaoa,chancellor2019domainwall,sawaya2022dqir,tabi2020coloring,glos2022qaoa,mcardle2019vibr}. 
\mtq~ was developed primarily for the specific task of preparing and compiling various problems into qubit representations; for other aspects of the compilation pipeline and for running the resulting qubit-based problems on quantum computers, we suggest the use of one of the many actively maintained quantum simulators \cite{qsharp2018,iqs2020,pennylane,cirq,quest2019,qiskit2019,qtorch,tnqvm2018,jet2022,gray2018quimb,tequila2021,projectq2018,qulacs2021,xacc2020} and compilers
\cite{qcor2020,scaffcc2014,pyzx2019,openql2021,projectq2018,cirq,qiskit2019,qsharp2018,xacc2020}.










\section{Program flow}

Here we provide an overview of the \mtq~ program flow, focusing primarily on the most useful front-end classes and subroutines. The code's dependencies are \ttt{scipy}, \ttt{sympy}, and \ttt{openfermion}; the main component used from the latter package is the convenient \ttt{QubitOperator} class.
Our primary goal was usability, and as such there are several potential improvements in terms of computational performance. For instance, the encoding procedure could be sped up by using linear algebra subroutines related to the Walsh-Hadamard transform \cite{fino1976walshhadam}; further, the code could be parallelized using common pythonic packages \cite{lam2015numba,dalcin2021mpi4py}. Notably, despite the omission of such optimizations, we have used the code to compile problems up to thousands of qubits and published results up to hundreds of qubits \cite{schmitz2021graph}, meaning that the current version may be used for analyses well beyond what existing quantum hardware is capable of simulating.



\subsection{Operators over $d$-level subsystems. }
Three of the fundamental front-end classes of \mtq~ are \ttt{dLevelSubsystem}, \ttt{compositeDLevels}, and \ttt{composite Operator} (the fourth, \ttt{qSymbOp}, is discussed below). The first of these is used to define a quantum particle (\textit{i.e.} ``subsystem''). For an operator to be eventually converted into a qubit representation, the parameters \ttt{d} (cardinality of the variable) and \ttt{enc} (the chosen encoding) must be specified. The following code defines a subsystem with a truncation of 8 levels.
\begin{lstlisting}[]
import mat2qubit as m2q
bosonic_mode_1 = m2q.dLevelSubsystem(d=8,enc="stdbinary")
bm1_pauli_operator = bosonic_mode_1.opToPauli("ad")
\end{lstlisting}
The third line converts the creation operator ($a^\dag \equiv$ \ttt{ad}) to a qubit representation.

Collections of subsystems are organized using the \ttt{compositeDLevels} class. However, the user will most often interact with its child class \ttt{compositeOperator}. Subsystems are added using \ttt{appendSubystem()}, and terms are added using \ttt{addHamTerm()}. The following example implements a two-site Hamiltonian using creation (\ttt{ad}) and annihilation (\ttt{a}) operators. 
%
\begin{lstlisting}[]
ss1 = m2q.dLevelSubsystem(4, "unary")
twoSiteOperator = m2q.compositeOperator()
twoSiteOperator.appendSubsystem(ss1) # Site 1
twoSiteOperator.appendSubsystem(ss1) # Site 2
twoSiteOperator.addHamTerm(4.2, [(1,"ad"),(0,"a")] )
twoSiteOperator.addHamTerm(4.2, [(0,"a"),(1,"ad")] )
twoSiteOperator.addHamTerm(0.5, "ident" ) # Identity
pauli_operator = twoSiteOperator.opToPauli()
\end{lstlisting}
The final line converts the operator to a qubit (Pauli) representation. There is no requirement that subsystems have the same size $d$; hence heterogeneous quantum systems are allowed. For instance, one may implement a spin-boson model using a collection of two-level particles and bosonic particles \cite{wall2016spinboson}. Nor do subsystems need to use same encoding; indeed, there may be cases when using different encodings for different subsystems leads to a compiled operator with fewer Pauli terms \cite{sawaya2020dlev}. 



\subsection{Compiling into qubit operators}

Once an operator has been defined, the goal is to implement the mapping of Equation \eqref{eq:general_mapping}. \mtq~ has built-in options to compile the higher-level multi-particle operator into a qubit operator. The final step of this procedure is to run the \texttt{opToPauli()} method in the last line of the previous section.
The built-in encodings are unary (also called one-hot), standard binary (SB), Gray code, and block unary. The latter is in fact a class of encodings, for which one must specify the local encoding and the block size $g$ \cite{sawaya2020dlev,sawaya2020connec,sawaya2022dqir}. 
For a $d$-level particle, unary requires $d$ qubits, Gray and SB require $\ceil{\log_2 d}$ qubits, and block unary requires $\ceil{\frac{d}{g}}\ceil{\log_2(g+1)}$, where $\ceil{\cdot}$ is the ceiling function. 

Note that each particle in the original space is mapped to its own disjoint set of qubits. In principle, more complex mappings are possible, for example different $d$-level particles could map to the same subset of qubits (perhaps in order to reduce space requirements), though the current code is not capable of such embeddings. 


Novel encodings beyond those mentioned above may be added. To do so, the user must define functions for both the integer-to-bit mapping and the \textit{bitmask subset}, which determines the set of qubits over which a particular matrix element operates. Previous work discusses this concept more thoroughly \cite{sawaya2020dlev,sawaya2022dqir}. Once these functions are defined they may be introduced by directly modifying \ttt{integer2bit.py}. Note that the current infrastructure is built to handle encoding parameters as well---for instance, parameters are required in order to fully define a block unary encoding, and other user-defined parametrized encodings are allowed.

\subsection{\ttt{mat2qubit}'s symbolic computer algebra system}

For ease of programmability, we have included a computer algebra system (CAS) that simplifies operator algebra. Our \ttt{qSymbOp} class allows for construction and algebraic manipulation of fully symbolic operators, where local operators, subsystem (particle) labels, and coefficients may all be defined symbolically. Allowing all three of these entities to be labelled symbolically allows for improved programmability and readability. For example, one may use symbolic coefficients and then pass the symbolic object to various functions that implement different numerical parameters. Further, the ability to label subspaces with arbitrary character strings can prevent confusion, for example clarifying when different particles are of different types; if one were to model a chain of spins each coupled to a bosonic bath, one might label spin subsystems with \ttt{\_sp0}, \ttt{\_sp1}, etc. and bosonic modes with \ttt{\_b0}, \ttt{\_b1}, etc.

The following is an example of the use of \ttt{qSymbOp}, with operators defined over subystems \ttt{A}, \ttt{B}, and \ttt{C}.
\begin{lstlisting}
symbolic_op1 = m2q.qSymbOp("k [n_A] ++ j [p_B] ++ r []")
symbolic_op2 = m2q.qSymbOp("k [q_A] ++ -r [p_C]")
# Product of operators
symb_product = symbolic_op1*symbolic_op2
# Subsitute numbers in for variables
symb_product.scalar_subs({'k':2.1, 'j':np.pi, 'r':3.}) 
\end{lstlisting}

Multiplication and addition operations may be implemented directly; the background data structures are appropriately updated to ensure there are no duplicate terms. Local operators are stored in order of multiplication as they may not commute. The final line of the above code fragment is used to convert the symbolic coefficients to numerical values, a step that must be done before the operator is compiled into a qubit representation. The final mapping is performed as follows, where three inputs must be given: the subsystem labels, $d$ values, and encodings.

\begin{lstlisting}
ssid_order = ['A','B','C']
ssid_to_d = [4,4,6]  # Heterogeneous d values  
ssid_to_enc = ['stdbinary','gray','stdbinary'] # Mixed mappings
dlev_obj = m2q.symbop_to_dlevcompositeop(symb_product,  ssid_order,ssid_to_d,ssid_to_enc)
pauli_operator = dlev_obj.opToPauli()
\end{lstlisting}
%




\subsection{Built-in and user-defined local operators}


There are many built-in operators, which are enumerated in the \ttt{builtInOps} list of \ttt{dLevelSystemEncodings.py}. Some are duplicate; for instance both \ttt{qhoPos} and \ttt{q} may be used for standard bosonic position operator, while \ttt{qhoMom} and \ttt{p} are both for the momentum operator. Also notable are \ttt{q2}, \ttt{q3}, etc., which denote powers $q^2$, $q^3$, etc. of the position operator, used especially in molecular vibrational Hamiltonians \cite{mcardle2019vibr,sawaya2021infrared,ollitrault2020ansatz}. \ttt{Sx}, \ttt{Sy}, and \ttt{Sz} are the standard spin-$s$ operators for the three Cartesian directions. 

Further, there are built-in operators that can effectively be used to implement the discrete quantum intermediate representation (DQIR) \cite{sawaya2022dqir}. The DQIR is an intermediate representation that was previously introduced in order to facilitate compilation and analysis of classical optimization problems defined over discrete variables. Two ``primitives'' of the DQIR are the \textit{indicator primitive} $\mathcal P\supr{k}_\a \equiv |k \ra\la k|_\a,$ and the \textit{transfer primitive} $\mathcal T^{(k \leftarrow l)}_\a \equiv |k \ra\la l|_\a$. They are implemented in \mtq~ as \ttt{Pr\{i\}\_\{a\}} and \ttt{k\{i\}b\{j\}\_\{a\}} , respectively. For instance,
\begin{lstlisting}
Pr3_ss
\end{lstlisting}
denotes the projector (indicator primitive) $|3 \ra\la 3|$ and 
\begin{lstlisting}
k2b3_ss
\end{lstlisting}
denotes the ket-bra (transfer primitive) $|k \ra\la l|$, both operating on subspace \ttt{ss}. The \ttt{examples/} directory of the source shows examples of these indicator primitives being used in problem instances of graph coloring, traveling salesperson, and machine scheduling. The transfer primitives may be used for example to define QAOA mixers within the DQIR \cite{sawaya2022dqir}. 

The user is not required to use built-in operator definitions; arbitrary local $d$-level operators may be used with either \ttt{qSymbOp} or \ttt{compositeOperator} objects, in the following manner. For the latter, one may directly insert an operator string that uses \ttt{numpy} arrays instead of character strings:
\begin{lstlisting}[]
import numpy as np
locop1 = np.random.rand(d,d)
locop2 = np.random.rand(d,d)
twoSiteOperator.addHamTerm(4.2, [(1,locop1),(0,locop2)] )
\end{lstlisting}
%


For user-defined local matrix operators in \ttt{qSymbOp}, one may use any alphanumeric character string as long as the \ttt{inpOpChars} argument is specified when converting to a qubit operator. In the following example the local opererators \ttt{W} and \ttt{V} are defined by the user.
\begin{lstlisting}
symbolic_op3 = m2q.qSymbOp("2.0 [W_A] ++ 3.0 [V_B] ++ r []")
ssids = ('A','B'); ds = dict(zip(ssid_order,[4,6])); encs = ('stdbinary','gray')
Wrepr = np.random.rand(4,4)
Vrepr = np.random.rand(6,6)
dlev_obj = m2q.symbop_to_dlevcompositeop(
    symb_product, ssids, ds, encs,
    inpOpChars={'W':Wrepr,'V':Vrepr}
    )
pauli_operator = dlev_obj.opToPauli()
\end{lstlisting}
%










\subsection{Distribution and examples}



The source code and installation instructions are available at \url{github.com/IntelLabs/mat2qubit}. The package is distributed under the Apache 2.0 license. Many examples are available in the repository, where we have implemented simple instances of the following problem types: Bose-Hubbard models, anharmonic molecular vibrations, interacting spin-$\frac{3}{2}$ and spin-$\frac{5}{2}$ models, random tensor trains from classical linear algebra, graph coloring, the traveling salesperson problem, and machine scheduling.




\section{Concluding remarks}

We have introduced \ttt{mat2qubit}, a lightweight python package for compiling quantum and classical problems into qubit representations. Specifically, it is intended for non-fermionic problems like bosonic, phononic, and spin-$s$ problems, as well as discrete optimization problems like scheduling, routing, and coloring. It may also be used to encode general matrix problems using various embeddings, as long as the matrix is first expressed as a sum of tensor trains. Theoretical details related to the code have been discussed previously \cite{sawaya2020dlev,sawaya2022dqir}. Before being open sourced, early versions of this package were used in numerical studies of several classes of Hamiltonian \cite{sawaya2020dlev,sawaya2022dqir,sawaya2019vibronic_jpcl,kyaw2021qcad,sawaya2021infrared,sawaya2020connec,schmitz2021graph}. The current release includes a computer algebra system for defining and manipulating quantum operators for which coefficients, local operators, and subsystem labels are all symbolic; this is intended to improve programmability, usability, and interpretability when constructing complex Hamiltonians. One main purpose of the code is to easily and rapidly implement many encodings for a given problem, in order to compare resource requirements. As quantum computers become more ubiquitous and their hardware characteristics more varied, such resource comparisons will be needed in order to determine the most efficient encodings for a given problem and a given set of hardware. In conclusion, we expect \ttt{mat2qubit} to be useful for compiling a wide variety of problems for quantum computers.


\section*{Acknowledgements}
I am grateful for useful discussions with Tim Menke, Stuart Hadfield, Gian Guerreschi, Shavindra Premaratne, Al\'{a}n Aspuru-Guzik, Thi-Ha Kyaw, and Albert Schmitz.







\bibliography{apssamp}

\end{document}